\documentstyle[aps,twocolumn]{revtex}
\begin{document}
\title{Fermi-liquid behaviour of the low-density 2D hole gas in GaAs/AlGaAs
heterostructure at large values of $r_s$.}
\author{Y.Y. Proskuryakov$^1$, A.K. Savchenko$^1$, S.S. Safonov$^1$, M. Pepper$^2$,
M.Y. Simmons$^2$, D.A. Ritchie$^2$}
\address{$^1$ School of Physics, University of Exeter, Stocker Road,\\
Exeter, EX4 4QL, U.K.\\
$^2$ Cavendish laboratory, University of Cambridge, Madingley Road,\\
Cambridge CB3 OHE, U.K.}
\maketitle

\begin{abstract}
We examine the validity of the Fermi-liquid description of the dilute 2D
hole gas in the crossover from 'metallic'-to-'insulating' behaviour of $\rho
(T)$. It has been established that, at $r_s$ as large as $29$, negative
magnetoresistance does exist and is well described by weak localisation. The
dephasing time extracted from the magnetoresistance is dominated by the $T^2$%
-term due to Landau scattering in the clean limit. The effect of hole-hole
interactions, however, is suppressed when compared with the theory for small 
$r_s$.
\end{abstract}


There has recently been much attention drawn to the unusual crossover in the
temperature dependence of resistivity with varying carrier concentration
from 'metallic' ($d\rho /dT>0$) to 'insulating' ($d\rho /dT<0)$ behaviour,
which has been seen in some low-density 2D systems \cite{many}. With
decreasing carrier density $p$ the ratio of the Coulomb to Fermi energy $%
r_s=U/E_F\propto m^{*}/p^{1/2}$ increases, so it was suggested that in these
systems the Fermi-liquid description is not valid and new approaches are
needed \cite{Bad-theories}. In \cite{Yoon} the insulating state at low hole
densities, $p\simeq 7.7\times 10^9$ cm$^{-2}$ and interaction parameter $%
r_s\simeq 35$ (with $m^{*}$ =0.37 $m_0$), was attributed to Wigner
crystallisation \cite{Tanatar}. At the same time, it was shown that even at $%
r_s\sim 10-14$ the 2D hole gas can manifest itself as a normal Fermi liquid,
as far as weak localisation (WL) and weak hole-hole interaction (HHI)
effects are concerned \cite{HamiltonWL,Senz} .

As conventional theories of weak localisation and electron interactions \cite
{AltshulerB,Hikami} are derived for $r_s\ll 1$, their applicability to $%
r_s\sim 10-14$ in \cite{HamiltonWL,Senz} is already a surprising fact. One
can argue though that in the structures studied the carrier density was not
low enough and mobility not high enough for the deviations from the Fermi
liquid description to be seen ($p\simeq 4.6\times 10^{10}$ cm$^{-2}$ and
peak mobility $\mu _p=2.5\times 10^5$cm$^2$V$^{-1}$s$^{-1}$ in \cite
{HamiltonWL}). In this work, we examine the existence of WL and HHI in a 2D
hole gas with lower density down to $p=1.17\times 10^{10}$ cm$^{-2}$ . In
our high mobility structures, with peak mobility $\mu _p=6.5\times 10^5$ cm$%
^2$V$^{-1}$s$^{-1}$, the experimental conditions approach those in \cite
{Yoon} where the Wigner crystal formation has been claimed. We show that
even at $r_s=29$ the strong Coulomb interaction does not affect the WL
description of the negative magnetoresistance \cite{WittmanSchmid,DmitrKoch}%
, although it suppresses the contributions to the phase-breaking time and
Hall coefficient of the weak hole interactions expected in a disordered
system with small $r_s$ .

The experiments have been performed on a high mobility heterostructure
formed on a (311)A GaAs substrate, where the 2D hole gas at the GaAs/AlGaAs
interface is separated from the Si-modulation doped layer by a 500\AA\
AlGaAs undoped spacer. A standard four-terminal low-frequency lock-in
technique has been used for resistivity measurements at temperatures down to
45 mK, with currents of 1-10 nA to avoid electron heating. The hole density $%
p$ is varied by the front gate voltage to provide the range of $r_s$ from $%
10 $ to $29$ (with effective mass $m^{*}$ taken as $0.38$ $m_0$\cite{Stormer}%
).

Fig. 1a shows a typical temperature dependence of the longitudinal
resistivity in our samples. The lower part of the plot, corresponding to
higher densities, has a 'metallic' behaviour with $d\rho /dT>0$. As the hole
density is decreased, $\rho (T)$ becomes nonmonotonic, and further
decreasing $p$ leads to an 'insulating' dependence with $d\rho /dT<0$. In
the 'metal' to 'insulator' crossover, where $r_s$ varies from $23$ to $29$,
we have observed negative perpendicular magnetoresistance, Fig. 1b, which
increases with lowering the hole density. It is natural to ascribe this
effect to WL which occurs due to quantum interference of elastically
scattered carriers on closed phase-coherent paths. However, great care
should be taken in analysing WL in high-mobility structures.

Firstly, the application of the conventional theory of WL is based on the
diffusion approximations and is restricted by the range of magnetic fields $%
B<B_{tr}$, where $B_{tr}=\hbar /4De\tau $ is the 'transport' magnetic field, 
$D$ is diffusion coefficient and $\tau $ is momentum relaxation time \cite
{Hikami}. Physically, this means that the magnetic length $L_B$ has to be
larger than the mean free path $l$. Within this approach, no negative
magnetoresistance is expected at $B>$ $B_{tr}$, when $L_B<l$ . In our high
mobility samples the value of $B_{tr}$ is very small, ranging from 0.003 to
0.08 T for the densities studied. At the same time, the NMR is observed up
to $\sim 0.2$ T, where Shubnikov-de Haas oscillations start. This means that
even at $B>B_{tr}$ there is a phase-breaking effect of magnetic field, which
acts on the trajectories which are still smaller than $L_B$ (and smaller
than $l$). The theory of WL in such a regime has been considered in several
papers \cite{WittmanSchmid,DmitrKoch}, although no experimental tests of the
theories have yet been performed.

Secondly, all theories of WL discuss the positive magnetoconductivity $%
\triangle \sigma _{xx}(B)=\delta \sigma _{xx}(T,B)-\delta \sigma _{xx}(T,0)$%
, which is due to the decrease of the negative correction $\delta \sigma
_{xx}(T,B)$ to the longitudinal classical (Drude) conductivity. Usually, the
phase-breaking effect of magnetic field is seen at small fields where its
effect on the Drude conductivity is negligible. In a high mobility system,
however, the phase-breaking effect will coexist with the magnetic field
dependence of the Drude conductivity itself:

\begin{equation}
\sigma _{xx}^D(B)=\frac{\sigma _0}{1+(\mu B)^2},  \label{Eq.1}
\end{equation}
where one cannot neglect the parameter $\mu B$. To analyse $\triangle \sigma
_{xx}(B)$ due to WL , the classical (negative) magnetoconductivity has to be
first subtracted.

Fig. 2 shows the total conductivity as a function of magnetic field,
obtained by inversion of the resistivity tensor , $\sigma
_{xx}^{tot}(B)=\left( 1/\rho _{xx}\right) /(1+\rho _{xy}^2/\rho _{xx}^2)$,
where $\rho _{xx}(B)$ and $\rho _{xy}(B)$ are measured simultaneously. The
dotted line is the classical expression Eq. (\ref{Eq.1}), with $\mu B=\rho
_{xy}(B)\sigma _0$. The zero field conductivity $\sigma _0$ is used as an
adjustable parameter to make the best fit in the higher field region where
WL is expected to be totally suppressed. For $\rho _{xy}(B)$ we use the
expression $\rho _{xy}=B/ep$, where the hole density $p$ is determined from
Shubnikov-de Haas oscillations in the higher field regime. The difference
between the solid and dotted lines then gives the WL correction $\delta
\sigma _{xx}(T,B)$, which decreases to zero at high fields where the
classical effect dominates. From this difference the zero-field value $%
\delta \sigma _{xx}(T,0)$ is then subtracted to obtain the required
dependence $\triangle \sigma _{xx}(B)$.

To analyse our experimental data, we use the WL theory \cite{WittmanSchmid}
developed beyond the diffusion approximation. It gives the
magnetoconductivity at an arbitrary ratio $B/B_{tr}$, provided $\tau <\tau
_\varphi $ is satisfied:

\begin{equation}
\Delta \sigma (B)=\frac{-e^2}{\pi h(1+\gamma )^2}\left[ \sum_{n=0}^N\left( 
\frac{b\cdot \psi _n^3(b)}{1+\gamma -\psi _n(b)}\right) -\ln \frac{1+\gamma }%
\gamma \right] ,  \label{eq2}
\end{equation}
where $\gamma =\tau /\tau _\varphi $, $\tau _\varphi $ is the dephasing
time, $b=\frac 1{(1+\gamma )^2}\frac B{B_{tr}}$, $\psi _n(b)=\int_0^\infty
d\xi \cdot e^{-\xi -b\xi ^2/4}L_n(b\xi ^2/2)$, and $L_n$ are the Laguerre
polynomials. In Fig. 3a we show representative data at different densities
in the middle of the temperature range studied, plotted against
dimensionless magnetic field $B/B_{tr}$ ($B_{tr}$ is found as $(4\pi \hbar
\mu \sigma _0/e^2)^{-1}$). Solid lines in Fig. 3 are obtained from Eq. (\ref
{eq2}), where $\gamma $ is used as an adjustable parameter. At lowest $p$
and $T$ the error in determining $\gamma $ is 10\%, but as the density
or/and temperature increases it steeply drops to 5\%.{\bf \ }The obtained $%
\gamma $-values range from 0.04 to 0.43 and satisfy the condition $\tau
<\tau _\varphi $.\ 

The{\bf \ }apparent agreement with WL theory suggests that, surprisingly,
even at $r_s\sim 23-29$ the Fermi-liquid description of the system remains
valid. The further evidence of this has been obtained from the analysis of
the temperature dependence $\tau _\varphi ^{-1}(T)$ of the dephasing rate.
Estimations show that the contribution to $\tau _\varphi ^{-1}(T)$ of
electron-phonon scattering \cite{Karpus} is negligible in the studied
temperature range. According to the Fermi-liquid theory \cite
{Reizer,FukuyamaAbr,AltshulerB}, the dephasing rate due to electron-electron
scattering is dominated either by a linear or quadratic term, dependent on
the parameter $\tau k_BT/\hbar $:

\begin{eqnarray}
\tau _\varphi ^{-1}(T) &=&\alpha \frac{(k_BT)^2}{\hbar E_F}\ln \left( \frac{%
4E_F}{k_BT}\right) ,\text{ when }k_BT\tau /\hbar \gg 1  \label{eq3} \\
\tau _\varphi ^{-1}(T) &=&\frac{k_BT}{2E_F\tau }\ln \left( \frac{2E_F\tau }%
\hbar \right) ,\text{ when }k_BT\tau /\hbar \ll 1
\end{eqnarray}
where $\alpha =\pi /8$ \cite{Reizer}. The quadratic term{\bf \ }in Eq. (\ref
{eq3}) is due to Landau-Baber scattering associated with collisions in a
clean Fermi-liquid with large momentum transfer, and the linear term in Eq.
(4) corresponds to particle-particle interactions with small energy transfer
in disordered conductors. In the experiment, the parameter $k_BT\tau /\hbar $
varies from 0.06 to 0.8 for the lowest studied density and from 0.1 to 0.9
for the highest density, so that we need to examine the applicability of
both expressions to the dependence $\tau _\varphi ^{-1}(T)$ extracted from
the analysis of the magnetoresistance data, Fig. 4a.

It is interesting to note that in the whole range of $p$, including the
lowest densities with 'insulating' dependence $\rho (T)$, we have seen
Shubnikov-de Haas oscillations. In the studied sample a shift of the
Shubnikov-de Haas minima was seen with increasing temperature from 45 mK to
600 mK, indicating a weak ($\sim 10\%$) increase of the hole density. Thus,
it was convenient to analyse the dephasing rate as the product $\tau
_\varphi ^{-1}\cdot p$, with density $p$ directly measured at each
temperature by the Shubnikov-de Haas effect. In this case Eqs. (3-4) are
re-written as 
\begin{eqnarray}
\tau _\varphi ^{-1}\cdot p &=&\frac{m^{*}}{\pi \hbar ^3}\left[ \alpha
k_B^2T^2\ln \left( \frac{4E_F}{k_BT}\right) \right]   \label{eq5} \\
\tau _\varphi ^{-1}\cdot p &=&\frac{m^{*}}{\pi \hbar ^3}\left[ k_BT\frac 
\hbar {2\tau }\ln \left( \frac{2E_F\tau }\hbar \right) \right] 
\end{eqnarray}
The experimental curves in Fig. 4a show two distinct features: a non-linear
form and a saturation at low temperatures. We have established that in the
entire range of hole densities and temperatures, the data are well described
by the quadratic term, Eq. \ref{eq5}, with a zero-temperature saturation
value $1/\tau _\varphi (T=0)=1/\tau _\varphi ^{sat}$ added to it. In the
analysis of $\tau _\varphi \left( T\right) $ we used values $p$, $E_F$ and $%
\tau $ experimentally determined at each $T$. Coefficient $\alpha $, found
as an adjustable parameter, agrees within 20\% accuracy with the value $\pi
/8$ for all hole densities, inset to Fig. 4a. At the same time, the data
show that the linear term in the dephasing rate is suppressed by more than
an order of magnitude compared with the value estimated using Eq. 6 \cite
{AltshulerB}. The expected contribution to the dephasing rate for the middle
density $p=1.3\times 10^{10}$ cm$^{-2}$ is shown in Fig. 4a as a solid line
(which is practically a straight line due to the weak ($\sim 10\%$)
temperature dependence of $\tau $ and $E_{F\text{ }}$).

We suggest two possible reasons for the suppression of the linear term.
According to the conventional theories \cite{AltshulerB,FukuyamaAbr}, the $T$%
-term originates from electron-electron scattering in systems with a
diffusive character of transport, when the interaction potential between
electrons is weak $(r_s<<1)$. The interactions in our system are strong. In
addition, in our high mobility system, the coherent paths of size $L_\varphi 
$ contain only a small number of scatterers ($\sim 10$), thus the diffusion
approximation may not be valid. The non-diffusive transport does not affect,
however, the $T^2$ term which is an intrinsic property of clean systems.

Let us briefly discuss the saturation of $\tau _\varphi ^{-1}$. The problem
of saturation of the dephasing rate at low $T$ has been known for many years 
\cite{AltshulerAleiner,Webb}, with several explanations of this effect
suggested. A characteristic feature of the saturation in our case is that it
becomes more pronounced with increasing density. Then a possible origin of
the saturation can be a nonequilibrium external noise which does not disturb
the temperature dependence of $\tau _\varphi ^{-1}$ in the Fermi liquid and
manifests itself simply as an additive to the dephasing rate \cite
{AltshulerAleiner}. According to \cite{AltshulerAleiner} the saturation due
to noise is $1/\tau _\varphi ^{sat}\simeq D^{1/5}(\Omega eE_{ac})^{2/5}$,
where $\Omega $ is the radiation frequency and $E_{ac}$ is amplitude of the
ac electric field. In Fig. 4b the value $1/\tau _\varphi ^{sat}$ is plotted
as a function of $D$ and shows agreement with the $D^{1/5}$ dependence.

In a conventional Fermi liquid, WL is usually accompanied by
electron-electron interaction effects, which are seen as quantum corrections
to the conductivity and Hall coefficient. At small $r_s$, the two
corrections are related as $\delta R_H(T)/R_H=-2\delta \sigma (T)/\sigma $ 
\cite{AltshulerB}. In \cite{HamiltonWL} it was argued that in {\em p}-GaAs
heterostructures the weak hole-hole interaction effect persists up to $%
r_s\sim 10-14$. Now we have measured the temperature dependence of the Hall
coefficient at much larger $r_s$. In Fig. 4c we plot the Hall coefficient as 
$\left( R_He\right) ^{-1}$for different temperatures in the range from 45 to
400 mK at the density $p=1.45\times 10^{\text{10}}$ cm$^{\text{-2}}$ (solid
squares). The decrease of $R_H$ with increasing $T$ appears to be of the
same order of magnitude as that estimated from theory \cite{AltshulerB}.
However, the observed shift of Shubnikov-de Haas minima has indicated an
increase of the hole density with temperature in this experiment (circles in
Fig. 4c), such that it agrees well, within 2\%, with the change in $R_H$ .
The inset in Fig.4c shows good agreement between the densities measured by
Hall and Shubnikov-de Haas effects at different $V_g$ for $T=45$ mK. One can
then conclude that the interaction effects in the Hall coefficient appear to
be much weaker than expected from the perturbation theory derived at small $%
r_s$. We believe that this suppression of the temperature dependence in the
Hall coefficient has the same origin as the absence of the linear term in
the dephasing rate, namely strong interactions at large $r_s$ and, possibly,
breakdown of the diffusion approximation.

To summarise, we have investigated the applicability of the Fermi-liquid
description to a high mobility, low density 2D hole gas with large $r_s$,
approaching the conditions of expected Wigner crystallisation. We have found
that the negative magnetoresistance in the crossover region from 'metal' to
'insulator' persists up to $r_s$ $\sim 29$, and is caused by weak
localisation. The dephasing rate $\tau _\varphi ^{-1}(T)$ is dominated by a $%
T^2$-contribution due to Landau-Baber scattering, which is a characteristic
property of a Fermi-liquid in the clean limit. At the same time, the linear
in $T$ -term in the scattering rate, due to scattering with small energy
transfer, is suppressed. Its decrease is accompanied by absence of the
temperature dependence in the Hall coefficient. This demonstrates directly
that conventional understanding of interaction effects, developed at $r_s\ll
1$, has to be modified for a high-mobility Fermi liquid with strong
interactions.

We are grateful to B.L.Altshuler and D.L.Maslov for stimulating discussions,
Harry Clark for help with fabricating the samples, EPSRC and ORS fund for
financial support.\\

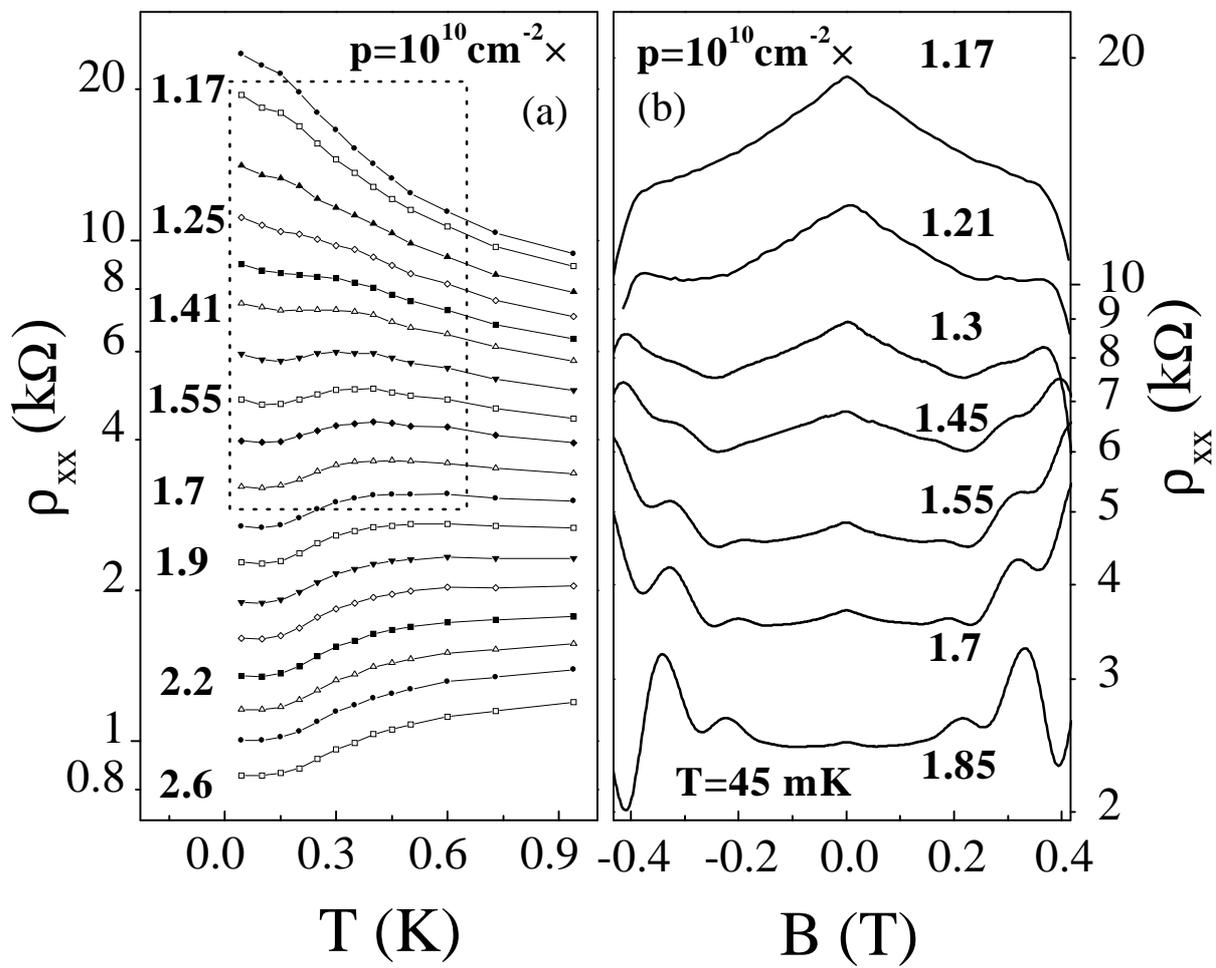
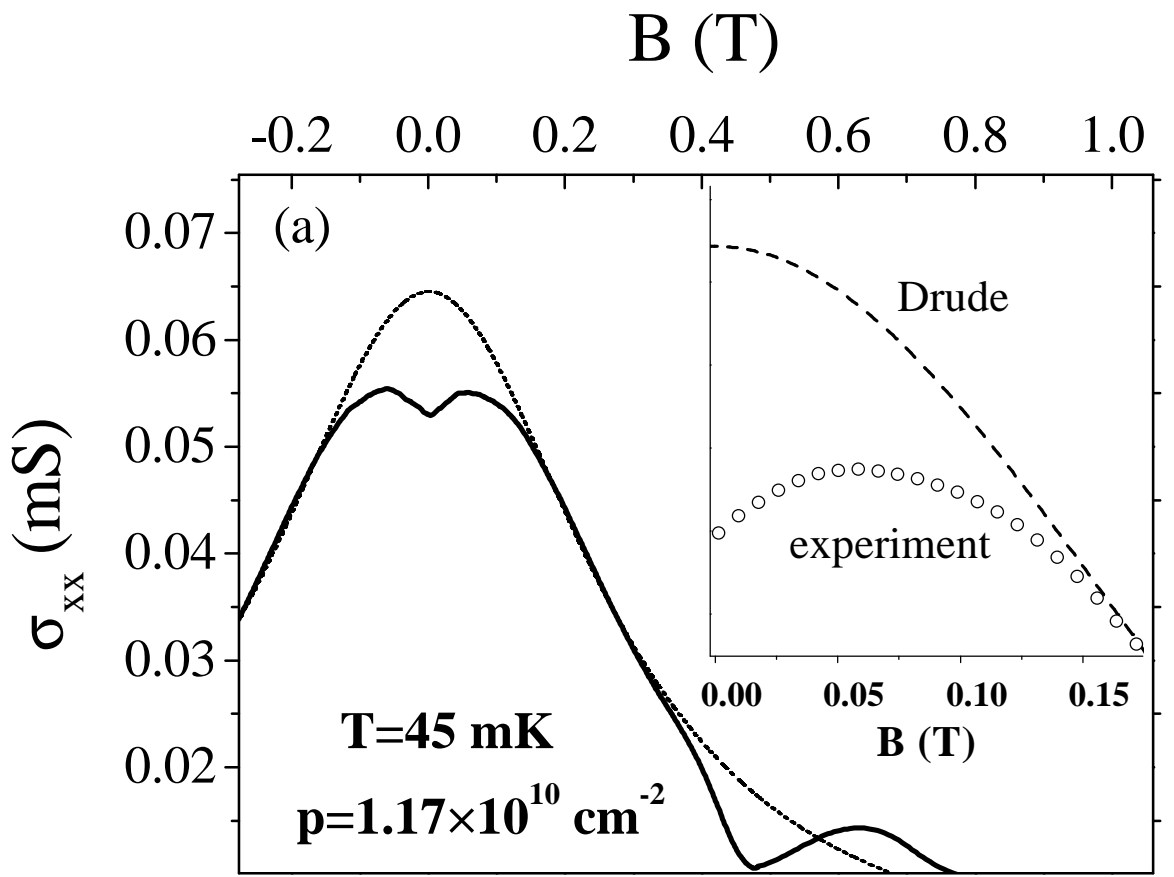
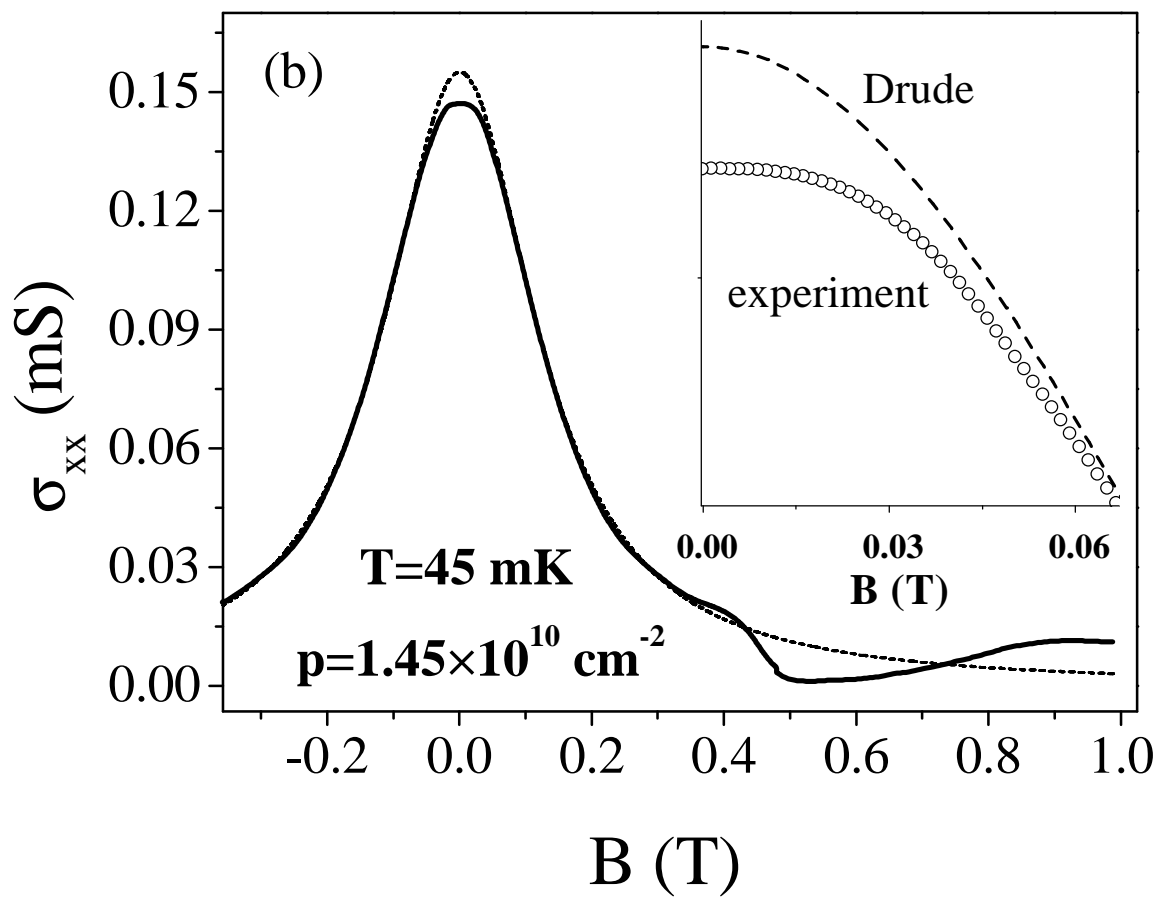
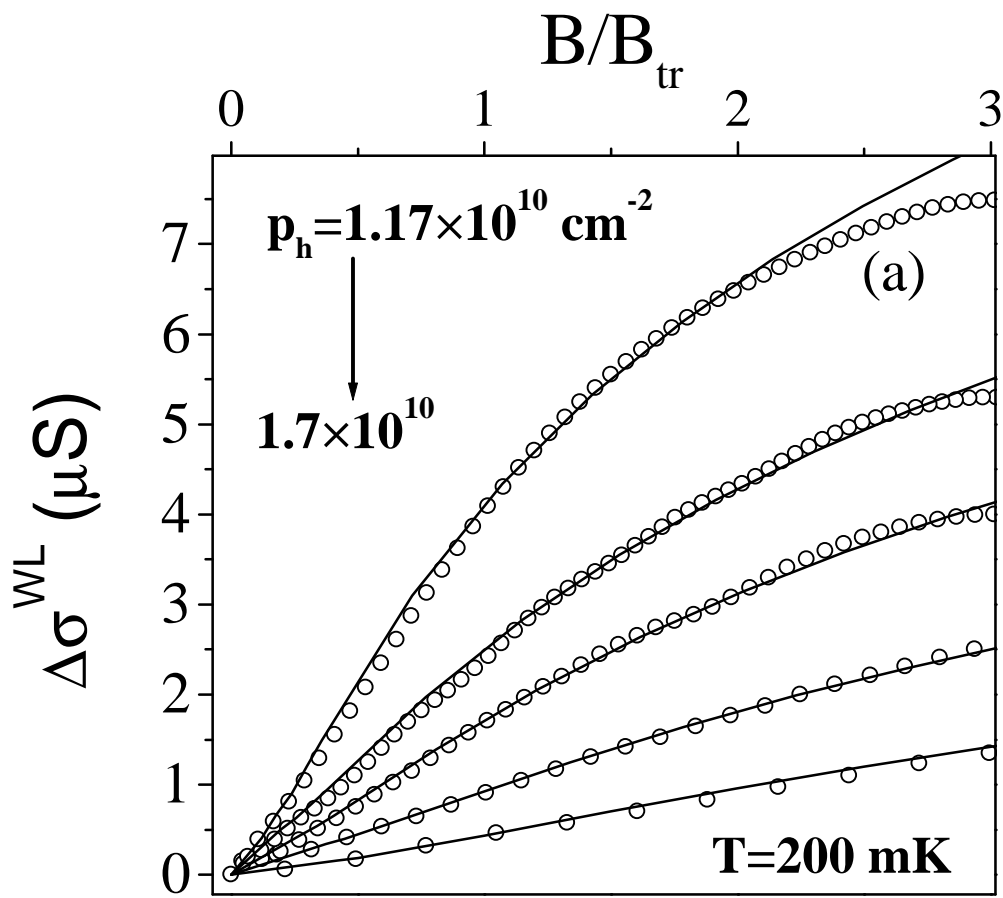
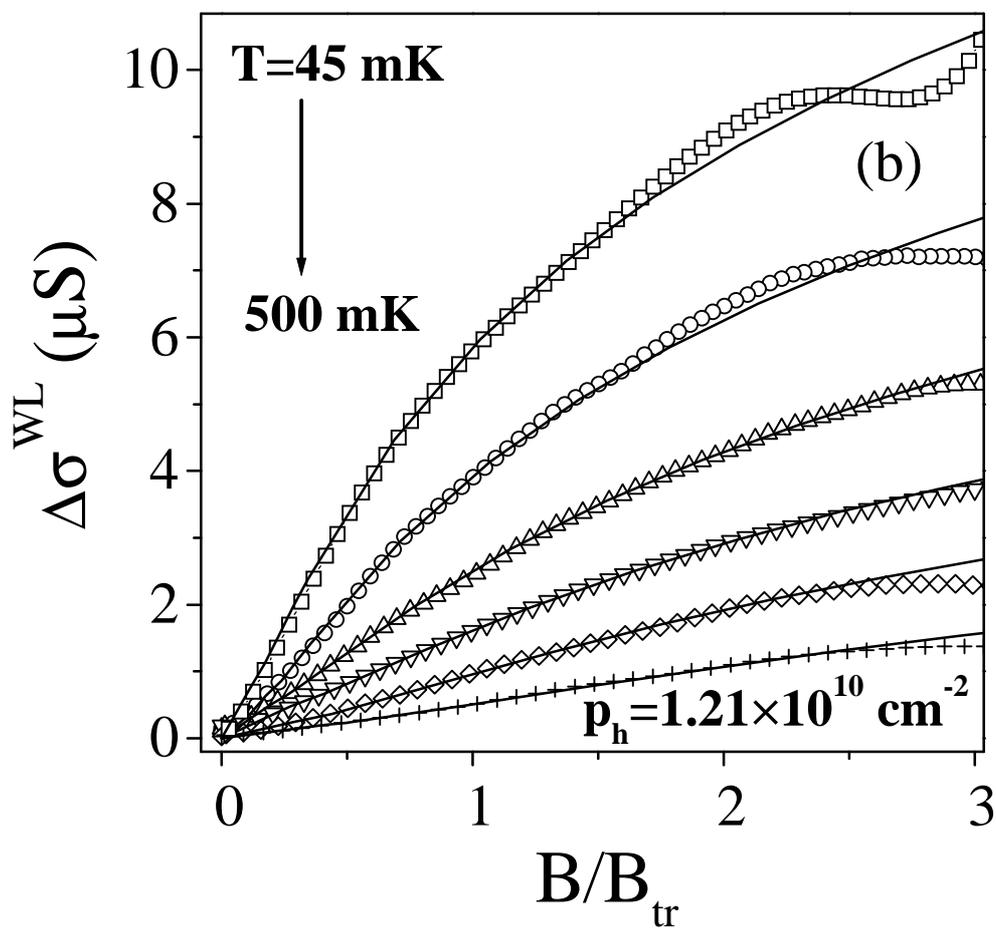
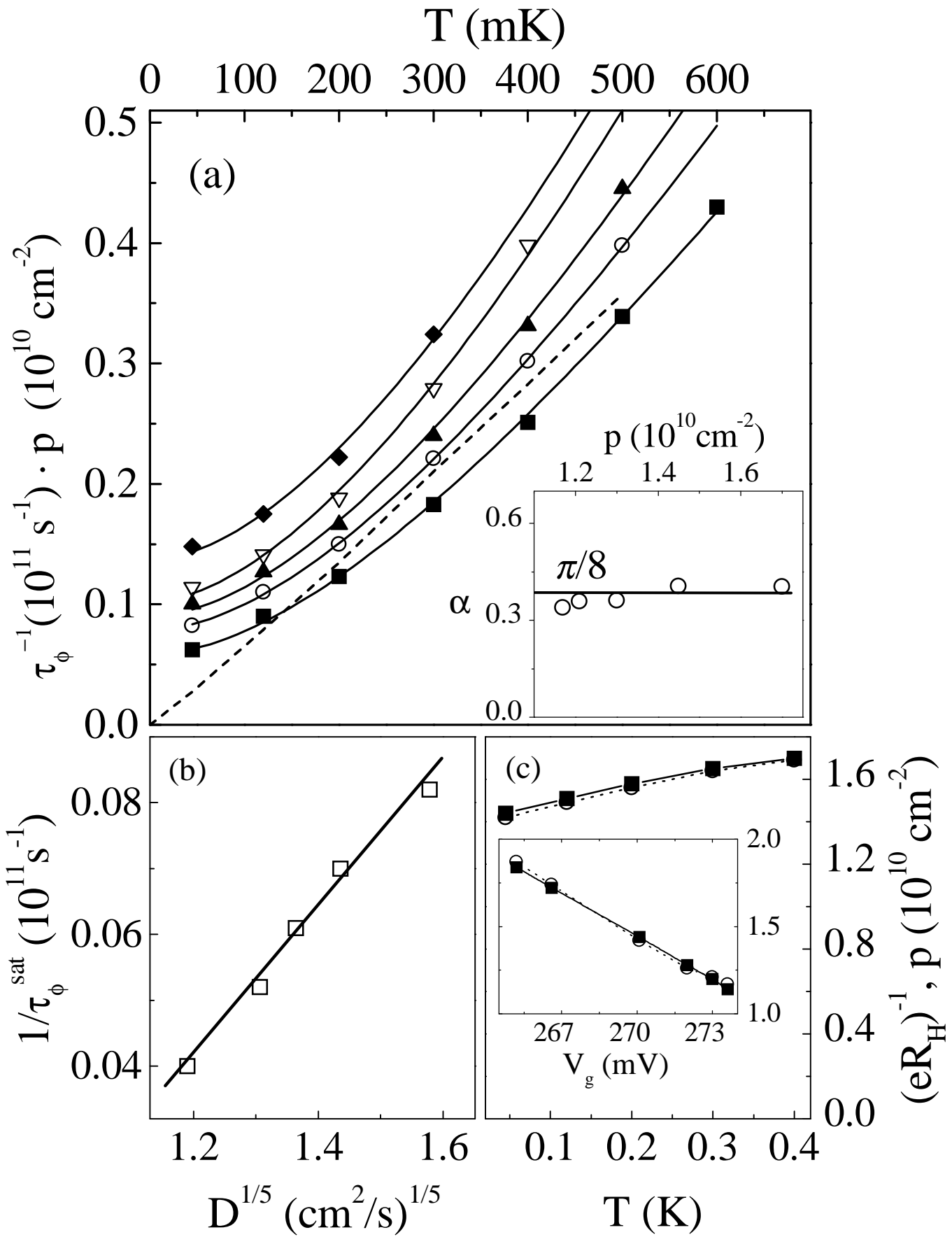
\end{document}